\documentstyle[psfig]{mn}

\newif\ifAMStwofonts

\ifoldfss
  \ifCUPmtlplainloaded \else
    \NewTextAlphabet{textbfit} {cmbxti10} {}
    \NewTextAlphabet{textbfss} {cmssbx10} {}
    \NewMathAlphabet{mathbfit} {cmbxti10} {} 
    \NewMathAlphabet{mathbfss} {cmssbx10} {} 
  \fi
  \ifAMStwofonts
    \ifCUPmtlplainloaded \else
      \NewSymbolFont{upmath} {eurm10}
      \NewSymbolFont{AMSa} {msam10}
      \NewMathSymbol{\upi}     {0}{upmath}{19}
      \NewMathSymbol{\umu}     {0}{upmath}{16}
      \NewMathSymbol{\upartial}{0}{upmath}{40}
      \NewMathSymbol{\leqslant}{3}{AMSa}{36}
      \NewMathSymbol{\geqslant}{3}{AMSa}{3E}

       \let\le=\leqslant
       \let\ge=\geqslant
    \fi
  \fi
\fi 

\ifnfssone
  \newmathalphabet{\mathit}
  \addtoversion{normal}{\mathit}{cmr}{\rm m}{it}
  \addtoversion{bold}{\mathit}{cmr}{bx}{it}
  \newmathalphabet{\mathbfit} 
  \addtoversion{normal}{\mathbfit}{cmr}{bx}{it}
  \addtoversion{bold}{\mathbfit}{cmr}{bx}{it}
  \newmathalphabet{\mathbfss} 
  \addtoversion{normal}{\mathbfss}{cmss}{bx}{n}
  \addtoversion{bold}{\mathbfss}{cmss}{bx}{n}
  \ifAMStwofonts
    \ifCUPmtlplainloaded \else
      \UseAMStwoboldmath
      \makeatletter
      \new@mathgroup\upmath@group
      \define@mathgroup\mv@normal\upmath@group{eur}{\rm m}{n}
      \define@mathgroup\mv@bold\upmath@group{eur}{b}{n}
      \edef\UPM{\hexnumber\upmath@group}
      \new@mathgroup\amsa@group
      \define@mathgroup\mv@normal\amsa@group{msa}{\rm m}{n}
      \define@mathgroup\mv@bold\amsa@group{msa}{\rm m}{n}
      \edef\AMSa{\hexnumber\amsa@group}
      \makeatother
      \mathchardef\upi="0\UPM19
      \mathchardef\umu="0\UPM16
      \mathchardef\upartial="0\UPM40
      \mathchardef\leqslant="3\AMSa36
      \mathchardef\geqslant="3\AMSa3E

       \let\le=\leqslant
       \let\ge=\geqslant
    \fi
  \fi
\fi 

\ifnfsstwo
  \DeclareMathAlphabet{\mathbfit}{OT1}{cmr}{bx}{it}
  \SetMathAlphabet\mathbfit{bold}{OT1}{cmr}{bx}{it}
  \DeclareMathAlphabet{\mathbfss}{OT1}{cmss}{bx}{n}
  \SetMathAlphabet\mathbfss{bold}{OT1}{cmss}{bx}{n}
  \ifAMStwofonts
    \ifCUPmtlplainloaded \else
      \DeclareSymbolFont{UPM}{U}{eur}{\rm m}{n}
      \SetSymbolFont{UPM}{bold}{U}{eur}{b}{n}
      \DeclareSymbolFont{AMSa}{U}{msa}{\rm m}{n}
      \DeclareMathSymbol{\upi}{0}{UPM}{"19}
      \DeclareMathSymbol{\umu}{0}{UPM}{"16}
      \DeclareMathSymbol{\upartial}{0}{UPM}{"40}
      \DeclareMathSymbol{\leqslant}{3}{AMSa}{"36}
      \DeclareMathSymbol{\geqslant}{3}{AMSa}{"3E}

       \let\le=\leqslant
       \let\ge=\geqslant
    \fi
  \fi
\fi 

\ifCUPmtlplainloaded \else
  \ifAMStwofonts \else 
    \def\upi{\pi}
    \def\umu{\mu}
    \def\upartial{\partial}
  \fi
\fi

\title[COSMOSOMAS]{COSMOSOMAS: A Circular Scanning Instrument to Map the Sky at Centimetric Wavelengths}

\author[J. E. Gallegos et al.]
       {J. E. Gallegos$^1$, J. F. Mac\'\i as-P\'erez$^{2}$, 
C. M. Guti\'errez$^1$, R. Rebolo$^{1,\,3}$, \and R. A. Watson$^{2}$,    R. J. Hoyland$^1$ and S. Fern\'andez-Cerezo$^1$\\
$^1$ Instituto de Astrof\'{\i}sica de Canarias,  E~38200 La Laguna, Tenerife,
Spain\\
$^2$ University of Manchester, Jodrell Bank Observatory, Macclesfield, Cheshire, SK 11, 9DL, UK\\
$^3$ Consejo Superior de Investigaciones Cient\'{\i}ficas, Spain
}

\pagerange{\pageref{firstpage}--\pageref{lastpage}}
\pubyear{2000}
\def\deg~{\ifmmode^\circ _\cdot\else$^\circ _ \cdot$\fi }    
\def\degg{\ifmmode^\circ \else$^\circ $\fi }

\begin{document}

\maketitle

\label{firstpage}

\begin{abstract}

We describe  the first instrument of a Cosmic Microwave Background experiment 
for mapping cosmological structures on medium angular scales
(the COSMOSOMAS experiment) and  diffuse Galactic emission. The
instrument is located at Teide Observatory (Tenerife) and is based on a
circular scanning sky strategy. It consists of a 1 Hz spinning flat mirror
directing the sky radiation into a 1.8 m off-axis paraboloidal antenna
which focuses it on to a cryogenically cooled HEMT-based receiver
operating in the frequency range 12--18 GHz. The signal is split
 by a set of three filters, allowing simultaneous observations at 13, 15 and 17
GHz, each with a 1 GHz bandpass. A 1\degg\--5\degg\ resolution  sky
map  complete in RA and covering  20\degg~in declination is obtained each
day at these
frequencies. The observations presented here correspond to the first
months of operation, which have provided  a map of ~9000 square degrees
on the sky centred at DEC = +31\degg\ with sensitivities of 140, 150
and 250 $\mu$K per beam area in the channels at 13, 15 and 17 GHz
respectively. We discuss the design and performance of the instrument, the atmospheric effects, the reliability of the data obtained and prospects of achieving a
sensitivity of 30 $\mu$K per beam in two years of operation.

\end{abstract}

\newpage

\section{Introduction}

In recent years experiments on angular scales from several degrees to
a few arc minutes have started to delineate the CMB power spectrum offering a unique approach to the study of conditions in the
early history of the Universe. Current observations at large and intermediate angular
scales constrain the level of normalization of the Sachs--Wolfe plateau
(Bennett et al.  1996; Guti\'errez et al. 2000), and have shown the
presence of the first Doppler peak (de Bernardis et al. 2000; Hanany et
al. 2000; Mauskopf et al. 2000; Halverson et al. 2001). A new generation of experiments is planned to cover
angular scales ranging from a few arc minutes to several degrees, these
include the {\it MAP\/} and {\it Planck\/} satellite missions. These experimental
efforts will potentially allow the determination of the main
cosmological parameters at a level of a few per cent. To achieve this, a
very accurate subtraction of foregrounds is needed. These foregrounds
include  synchrotron, free--free and dust emission. Recently, a fourth
component has been identified (Kogut et al.  1996a,b) in the
analysis of the {\it COBE\/} DMR data. The nature of this component is
controversial and has been  proposed to be free--free (Mukherjee et al.
2001) or spinning dust (de Oliveira-Costa et al. 1999, 2000). Progress
in the study of this elusive component will require the existence of
reliable maps at frequencies in the range between 10 and 20 GHz where
free--free emission is one of the dominating processes and spinning
dust (Draine \& Lazarian 1998) may exhibit a turn-over in its spectrum.

The goal of the COSMOSOMAS experiment
presented here is to map the cosmic microwave
background and  Galactic diffuse emission with mean sensitivities of
30 $\mu$K per beam ($\sim 1^{\circ}$) in an area covering $\sim 25$ \%
of the sky. This will allow a measurement of CMB fluctuations in the
angular regions of the Sachs--Wolfe plateau and the first acoustic peaks. The
experiment is based on a circular
 scanning strategy and consists of two ground-based total power receivers working
at central frequencies of 10 and 15 GHz respectively with a beam size of 1\degg. Observations with  a prototype of the COSMOSOMAS
experiment operating at 10 GHz (Gallegos et al. 1998)  proved that
the atmospheric conditions at Teide Observatory were suitable for
CMB experiments based on a circular scanning strategy. Although the
instrumental set-up is similar for both experiments, this paper
concentrates on a description of the 15 GHz instrument (which hereafter we
shall refer to as COSMO15), an analysis of its performance and a
presentation of its first results.

After describing the instrumental set-up (Section 2), the observations
and data analysis
(Section 3), we analyse the instrument's performance (Section 4) and present
maps of the sky obtained with COSMO15 (Section 5).

\section{The instrumental set-up}

\subsection{Optical System}

The COSMO15 instrument consists of a 2.5 m flat circular mirror whose
normal rotates about the central axis of spin
at a rate of 1 Hz and reflects the radiation  from the sky
into a parabolic dish. The spin axis can be tilted to change the
region of sky observed. The mirror is canted at 5$^\circ$ relative to its
spin axis to generate a circular path on the sky with a diameter
of 20$^\circ$. This geometrical configuration combined with the Earth's
rotation allows for complete coverage of a full declination strip 
once per sidereal day. The declination range covered is given by
$$\delta_{\circ} + \beta - 2 \alpha - 2 \gamma \le \delta \le
\delta_{\circ} + \beta - 2 \alpha + 2\gamma,
$$
\noindent
where $\delta_{\circ}$ is the site's zenith declination, $\alpha$ is
the inclination of the support of the mirror, $\beta$ is the angle of
the parabolic dish beam with respect to the zenith and $\gamma$ is the mirror's tilt angle. The
radiometer beam is formed by a cooled corrugated feed horn that
underilluminates an ambient-temperature 1.8 m paraboloid, which in turn
illuminates the precessing mirror. This offset parabolic reflector with
a corrugated feed has minimal blockage, approximately equal $E/H$-plane
beam widths, and relatively low sidelobe response.  These properties are
ideal for CMB observations.  All  the beam-forming optics are inside a fixed
aluminium ground screen. Figure 1 shows the optical and mechanical configuration of the
instrument. 

\begin{figure}
\vspace{8cm}
\caption{{\it Top:\/} Scanning strategy of the experiment. {\it Bottom:\/} Mechanical and optical configuration of the experiment.}
\label{Fig:1}
\end{figure}

The spinning mirror is driven by an AC motor at a
constant speed of $\sim$ 60 rpm. The mirror has a tilting base that
enables the elevation of the instrument to be changed. The optical set-up points north
with a  range of possible declinations between 10\degg~and 60\degg. 
The spin axis of the mirror assembly has an optical encoder which generates one
 pulse per revolution to
control the real speed and to serve as a physical direction reference for
the instrument.

\subsection{Receivers}

The total-power receiver (see Figure 2)  at the focal point of a
paraboloidal antenna  comprises a  cryostat cooled to 20 K, which
houses an HEMT amplifier (first amplification stage), 
a single feed horn and the calibration source. 
There is a second amplification stage  at ambient temperature,
together with a band splitter, filters  and detectors before
the voltage to frequency (VTF) converters. The sky signal enters the receiver
through an 8 cm diameter and 3 mm thick polypropylene window. Two
overlapping aluminium baffles define the entrance aperture. One is
anchored to the $\sim $ 80 K stage, the other at ambient temperature.
Strips of aluminized mylar electrically connect the top of the feed to
the warm baffle and supress RF interference.

\begin{figure}
\centerline{\psfig{figure=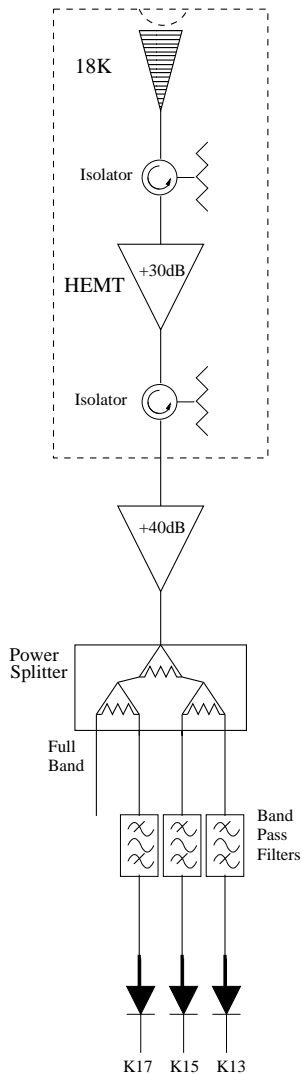,angle=0,width=4cm}}
\caption{COSMO15 receiver layout. Radiation enters
through the vacuum window and is collected by a cooled corrugated
feed.  The signal is amplified by a  broad-band HEMT amplifier
($\Delta\nu=6$ GHz), separated into three frequency bands and
square-law detectors.}   
\label{Fig:2}
\end{figure}

The beam is formed by a conical corrugated scalar feed. This feed is
matched with an electroformed adiabatic round-to-square transition to a
semirigid waveguide. The signal is fed directly into one K-band
Berkshire Technologies  Inc. amplifiers (K-15.0-25H). This device has a
6 GHz bandwidth ($\sim$ 12.0--18.0 GHz) and a +28 dB (min) RF gain.  The
noise temperature across the bandwidth for this amplifier is 12 K (max), when cooled
to $\sim$ 20\,K. The amplifier has isolators at the input and output to
prevent reflections.  The amplified signal is connected to the
back-end amplifiers (warm stage) through stainless-steel waveguides. The
RF signal is split into three 1 GHz bands centred at 12.5, 14.5 and
16.5 GHz (hereafter we will refer to these sub-bands as the 13, 15 and 17
GHz channels). Figure 3 shows the spectral response of these three channels.

\begin{figure}
\centerline{\psfig{figure=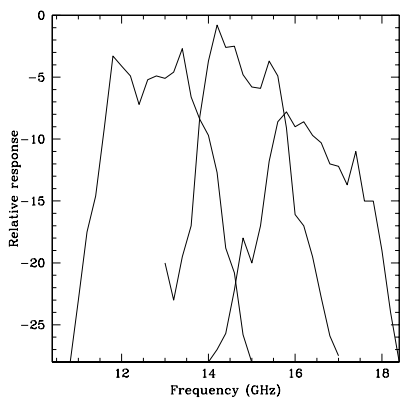,angle=0,width=5.5cm}}
\caption{Relative spectral responses of the 13, 15 and
17 GHz channels.}
\label{Fig:3}
\end{figure}

The cryogenic components are cooled by an APD Joule--Thomson
refrigerator.  All mechanical and electrical connections from the cold
stage to ambient temperature are thermally anchored to the refrigerator
$\sim$ 80\,K stage.  With an APD compressor, the cold stage runs at
$\sim$ 20\,K. The amplifiers and the corrugated horn are anchored to
this stage.  The room-temperature receiver box is rigidly mounted on
the cryostat base. The diode sensitivity and amplifier gain are a
function of temperature and accordingly the temperature of the
receiver enclosure is regulated for field operation.  The RF
components are fitted with insulation to prevent convective cooling and
are thermally anchored to a common aluminium mounting plate.

A thermally controlled noise diode radiates a calibration
 signal of $\sim$ 2.0 K directly into 
the corrugated horn for 1 second before and after 
each 30 seconds of observations. These calibration data are then
used to correct the gain fluctuations of the amplifiers. The
calibration data are then obtained by subtracting the signals with 
{\it cal-off} from the those with {\it cal-on}.

\subsection{Data acquisition system}

The output voltage signal from the three channel detectors are read with VTF converters which
are connected to the detectors with shielded pairs to minimize earthing problems. Along with
the receiver channels, the spin encoder is read once per revolution, giving a total of four
digital signals fed into the VTF converters.
The output signals from the VTF converters are read by a counter card in a  PC computer 
situated in an adjacent building. This computer also controls the square wave used to drive 
the calibration diode on and off. The computer is connected to the VTF converters and the calibration 
diode via a fast link which consists of two TTL-ECL converters, one at each of the connecting points.

The data are sampled each 4000 $\mu$s with a blanking time (no recording) of $400 \mu$s. This is
equivalent to have three samples per beam and a total of $\sim 220$ samples per turn of the mirror. 
The computer temporally stores 30 seconds of data (about 30 turns of the mirror).  
For each turn of the mirror a lock-in at multiples of the spin cycle is performed by decomposition
into a Fourier series. The first 106 Fourier coefficients (harmonics hereafter) which correspond
to 212 samples per turn are kept and stacked across the 30 second period.  
The stacked harmonics are then stored in a FITS file. This procedure does not only saves hard-disk space on 
the PC but also accelerates the data reduction, because the 212 samples represent fixed positions
on the sky. This is possible because the change in RA caused by the earth rotation in 30 seconds
 $\sim 0^{\circ}.12$ is 
negligible with respect to the beam-width $\sim 1^{\circ}$.

\section{Observations, data processing and astronomical calibration}

The instrument is installed at the Teide Observatory, which has been
shown to be a good site for centimetre and millimetre CMB
observations (Davies et al. 1996; Dicker et al. 1999). The instrument
started operation on 1999 September 1  and has remained
operational until 2000 September, with interruptions due to instrumental tests,
adverse atmospheric conditions and technical failure. During this period we have observed three
overlapping regions of the sky between 16\degg~and 46\degg~in
declination. About 100 days of data were obtained, of which we have
selected  the best $\sim$ 50 days in terms of atmospheric  quality in
order to generate a first set of maps. Most of the observation time was devoted
to the two high declination regions $22^{\circ}-46^{\circ}$ while at the low declination
region $16^{\circ}-36^{\circ}$ only 8 days of good observations were obtained.

\subsection{Scan reconstruction, offset removal and map making}

The individual scans are reconstructed from the stored harmonics via an FFT
 using IDL routines on a
workstation. 
A scan consists of 212 points which
represent fixed positions in hour angle (HA) and declination (DEC). The
top diagram of Figure 4 represents a typical COSMOSOMAS scan containing
the point source Tau A. The main modulation observed is  due mostly to the
change in air masses and ground pickup as the instrument scans a circle on the sky.
To reduce the dominant ground pick-up effect, we have covered the ground surrounding 
the instrument with aluminum
plates and have increased the size of the spinning mirror that is
currently heavily under-illuminated. Data affected by the Sun or Moon were removed.

\begin{figure}
\centerline{\psfig{figure=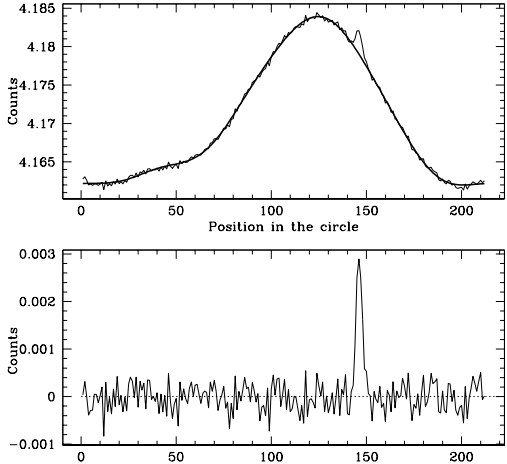,angle=0,width=7cm}}
\caption{Tau A crossing in a  30 sec scan. The modulation due
to atmospheric and ground spillover contribution is corrected by
fitting a sine--cosine function to the cycle (heavy line).}
\label{Fig:4}
\end{figure}

To remove these spurious signals, we use Fourier series to
subtract low Fourier components from the data. Firstly, we reduce
the effect of changes in the gain of the instrument by dividing the data
by the calibration signal. Secondly, we fit each of the scans to
a Fourier series of 11 coefficients, which means five sines and cosines plus a
total power level and subtract the fit from the data.  Therefore, angular
scales larger than 5 degrees are removed from the data, limiting the angular resolution
range to $1^{\circ}-5^{\circ}$, but most of the atmospheric contribution is also removed.
This procedure works very well
but needs to be repeated iteratively so that strong astronomical
features such as the Galaxy are preserved and no extra baseline is
introduced around them. At each fitting step a re-weighting of the data
is performed so that data three sigma away from the best fit are 
zero-weighted for the next step. After three or four iterations,
contributions from strong astronomical sources are weighted to zero and
the atmospheric shape is accurately reproduced by the fit. The upper
diagram of Figure 4 also displays such a fit to the scan, while the
lower diagram shows the residuals after baseline subtraction. 

Long drift baselines possibly due to changes in atmospheric conditions throughout the day
are still present in the data. To reduce these, we perform a second
fit to  the data. Each of the positions in the scans throughout the day
are fitted to Fourier series of seven coefficients and the fit is
subtracted from the data.  
This fitting procedure only removes from the data features at angular
scales larger than $20^{\circ}$ and therefore it does not affect the
$1^{\circ}-5^{\circ}$ structure we are able to observe.
Note that in this case we also perform an iterative
procedure to reduce the effect of Galaxy on the baseline fit. 
Once the fitting procedure is finished the clean scans are saved as IDL FITS
files along with the Julian date.

For each
clean scan we reproduce the instrumental pointing on the sky and derive
out RA and DEC for each position in the scan. A simple projection
scheme is used so that each RA and DEC position is converted into
pixel positions in the map. For each pixel in the map we calculate
an average contribution from all the scan positions lying within that
pixel. The mean temperature value and dispersion are calculated for each pixel. Points
within the pixel three sigma away from the mean value are excluded from
the final result and the map-making process is repeated iteratively.
The final map is composed of a mean value map, an error map and a
number of points per pixel map that are stored in a single IDL FITS
file. Pixels of $1/3\times 1/3$ degrees in RA and DEC are used because
they sample properly the beam response and minimize the noise
contribution per beam area.

The first region observed covers the range from $16^{\circ}$ to
$36^{\circ}$ in declination and the second one covers the range from
$25^{\circ}$ to $45^{\circ}$. Both have complete RA coverage. The
latter region is of special interest because it overlaps that observed
by the radiometers of the Tenerife Experiment (Guti\'errez et al.
2000) and the 33 GHz interferometer (Dicker et al. 1999, Harrison
et al. 2000) and will allow for a future comparison.

\subsection{Astronomical calibration}

Our primary calibration sources are the supernova remnant Tau A for the
low declination observations and Cyg A for the higher declinations. We
measure the beam of the instrument using these two sources.  Each beam
was fitted as an elliptical Gaussian to the main lobe. The sidelobes are  40 dB below the main beam. 
Figure~5 shows the
observations of  Tau A, and the two-dimensional fit of the beam.
From this analysis we conclude that the main beam is described by a
circular Gaussian with  FWHMs of 1.08$^\circ \pm 0.07^\circ$ at 13 GHz, 1.04$^\circ \pm 0.07^\circ$ at 15 GHz and 0.94$^\circ \pm
0.05^\circ$ at 17 GHz. The flux densities were estimated from the
compilation by Baars et al.  (1977). The resulting spectral fit for Tau A is
$\log ({\rm S_\nu /Jy})= (3.915\pm 0.031)- (0.299\pm
0.009)\log(\nu/{\rm MHz})$
in the 1 to 35\,GHz range and for Cyg A is
$\log ({\rm S}_\nu /{\rm Jy})= (7.161\pm 0.053) - (1.244\pm
0.014)\log(\nu/{\rm MHz})$
in the 2 to 31\,GHz frequency range.  Using the beam parameters
determined from these sources, we predict the antenna temperature for
Tau A and Cyg A, which are listed in Table \ref{calibrationflux} and these
values are used for the determination of the temperature scale.

\begin{figure}
\centerline{\psfig{figure=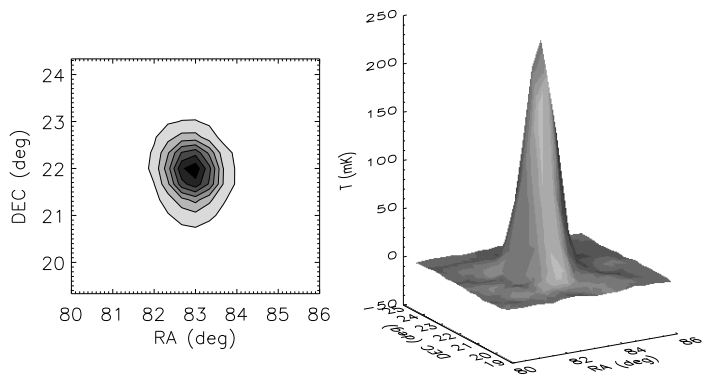,angle=0,width=11cm}}
\caption{ The reconstruction of  Tau A.}
\label{Fig:5}
\end{figure}

\begin{table}
\caption{Antenna temperatures for calibration sources}\label{Calibration} 
\begin{center}
\begin{tabular}{|c|c|c|} \hline
$\nu$ (GHz)  &  Tau A (mK)  &  Cyg A (mK) \\
\hline
13.0 & 246.7 & 52.6 \\
\\
15.0 & 190.0 & 36.2 \\
\\
17.0 & 173.3 & 29.1 \\
\\
\end{tabular}
\end{center}
\label{calibrationflux}
\end{table}

\section{Instrument Performance}

Figure 6 shows the power spectrum of a two-hour period of raw data
taken on 1999 November 22. We can observe three main features
in the plot. The ${1}/{f}$ noise component is seen at low frequencies ($< 0.3$ Hz) and the Gaussian
noise at high frequencies ($ > 4$ Hz). The knee frequency, defined as the frequency
at which the  Gaussian  and  ${1}/{f}$ noise have equal power, is
$\sim 2.7$ Hz. We also observe an intermediate regime (0.5 - 2.5 Hz) which
correspond to the atmospheric and ground pickup contributions which are superimposed
to the ${1}/{f}$ noise.

\begin{figure}
\centerline{\psfig{figure=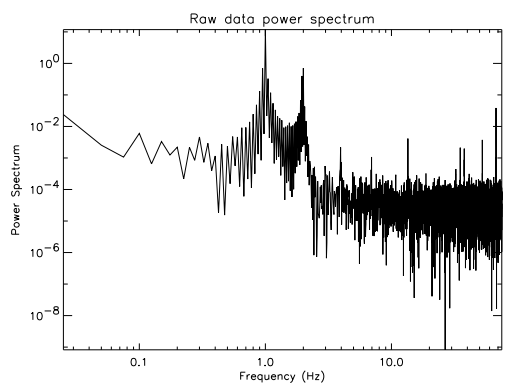,angle=+90,width=7cm}}
\caption{Power spectra of the raw data showing the $1/f$ noise.}
\label{Fig:6}
\end{figure}

The peaks at 1 Hz and 2 Hz are caused by the variation of the atmospheric emission and
the ground pickup with zenith angle for a turn of the mirror. We also observe peaks at 3, 4 
and 8 Hz which are the high frequency components of this variation. These features are synchronous
with the spinning frequency (1 Hz) of the mirror and stable between scans. Therefore they
cause the modulation discussed in section 3.1.
The dispersion observed between the 1 and 2 Hz peaks is mainly due to changes in the atmospheric
conditions. In the case of a non-spinning instrument this would show up as a noisy bump
in the spectrum. 

Finally, a 50 Hz peak is present and it is due to contamination from the mains. Extreme care
has been taken to reduce its contribution to a minimum; however complete elimination has not
been achieved. Nevertheless the 50Hz contribution is filtered out by software.

The sensitivity of the instrument depends basically on the efficiency
of the optical configuration and the system temperature.  We determined
the system temperature using standard hot and cold load calibration
procedures. We used a room-temperature absorber as a hot load and the
sky as a cold load and measured  the output power with a power meter
connected just before each channel detector. Table
\ref{cosmosystemtemp} contains the system temperature and the expected
r.m.s. noise per pixel $\sim 1.2$ mK in the 13 GHz, 15 GHz and 17 GHz
channels for a typical day of observation (2000 June 11).

\begin{table} 
\label{tabtempsys}
\caption{Theoretical sensitivity of COSMO15}
\begin{tabular}[t]{|c|c|c|}
\hline
 Frequency (GHz) & $T_{\rm sys} {\rm (K)}$	& ${\Delta T}_{\rm pixel} {\rm (mK)}$ \\
\hline
\\
 13			& 38.5			& 1.18			  \\
\\
 15			& 33.5		        & 1.17			  \\
\\
 17			& 37.9			& 1.18			  \\
\\
\end{tabular}  
\label{cosmosystemtemp}
\end{table}

The optical efficiency was estimated as follows. COSMO15 has a
calibration diode to monitor the receiver fluctuations.  The
temperature of this diode at each COSMO15 frequency can be measured if
the system temperature is known by applying a hot and cold load method
to diode on and off. The calibration factor measured this way does not
depend on the optical system of the telescope. However, the
astronomical calibration factor does depend on the characteristics of
the whole system. One can compare both factors and deduce the
efficiency of the system. For a 100 \% efficient system the values
should be similar to within the noise. Otherwise, the astronomical
calibration factor should be larger (i.e. the observed signal
smaller).  The ratio CAL$_{\rm m}$/CAL$_{\rm a}$ where CAL$_{\rm m}$ 
is the measured
calibration factor and CAL$_{\rm a}$ is the astronomical calibration factor, is
a direct measurement of the efficiency of the system. Table
\ref{tableefficiency} shows the measured and astronomical calibration factor
and the efficiency at the 13, 15 and 17 GHz channels for the data taken
on 2000 June 11.

\begin{table}
\caption{Calibration and efficiency of each channel}
\begin{tabular}[t]{|c|c|c|c|}
\hline
 Channel (GHz) &$\rm CAL_{\rm m} {\rm (mK)}$	& $\rm CAL_{\rm a} {\rm (mK)}$	& Efficiency  \\
\hline
\\
 13			&8150		& 12036			& 68 \%	   \\		
\\
 15			&7770		& 11019			& 70 \%   \\
\\
 17			&2920		& 3989			& 73 \%    \\
\\
\end{tabular}  
\label{tableefficiency}
\end{table}

We have repeated the above calculations on independent days and found
efficiency factors of 68, 69, and  74 \% at 13, 15 and 17 GHz
respectively. The noise figures calculated from the maps are larger
than expected from the system temperature of the instrument but not
fully compatible with the efficiency factors measured. Two extra factors
should be taken into account: error in the mapping process (mainly
arising from difficulties in the atmospheric destriping) and
extra noise coming from other unidentified contributions. Comparing
the expected noise and the noise calculated from single scans to the
noise found in the maps, we find that at 13 and 15 GHz the error in the
mapping process dominates. Table \ref{tablenoise} shows the noise
estimates calculated from individual scans and from the maps at 13, 15
and 17 GHz for 2000 June 11.

\begin{table}
\caption{Sensitivities in one day of observation}
\begin{tabular}[t]{|c|c|c|}
\hline
 Channel (GHz)	& ${\Delta T}_{\rm pixel}^{\rm scans} {\rm (mK)}$ & ${\Delta T}_{\rm pixel}^{\rm map} {\rm (mK)}$ \\
\hline
\\
 13			& 2.56		& 2.80	        \\
\\
 15			& 2.43		& 2.80		 \\
\\
 17			& 3.72		& 4.00		 \\
\\
\end{tabular}  
\label{tablenoise}
\end{table}

We have checked for the presence of systematic residuals in the
daily data and conclude that after subtraction of the  above mentioned offsets
the daily data are dominated by random noise. The noise per beam area $\sigma$
in the stacked maps is reduced as $\frac{1}{\sqrt{N}}$ where $N$ is the number of
stacked maps (Figure 7).

\begin{figure}
\centerline{\psfig{figure=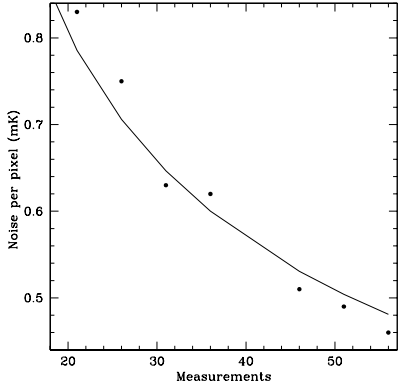,angle=0,width=5.5cm}}
\caption{The noise as a function of integration time.}
\label{Fig:7}
\end{figure}

We conclude that most of the increase of the noise in the maps at 13
and 15 GHz compared with the expected noise from the system temperature
is due to the inefficiency of the optical system. However, at 17 GHz we
have found another cause for the extra noise. Most
of the microwave components of the system have a cut-off in their
spectral behaviour for frequencies higher than 17 GHz, which could
produce a narrower effective bandwidth in the 17 GHz channel.

\section{The first COSMOSOMAS maps}

The final maps at each frequency resulting from observations conducted
up to 2000 June are presented in Figure~8. These comprise the
combination of 56, 49 and 46 good days of data at 13, 15 and 17 GHz,
respectively, obtained at the low- and the high-declination set-up. The
total region observed comprises a full band in RA and from 16\degg~to
46\degg~in declination, hence covering about 9000 square degrees. The
noise levels per beam area are ~140, 150 and 250 $\mu$K at 13, 15 and
17 GHz respectively. These sensitivities correspond to the high
galactic latitude part of the maps. The lower declinations are noisier
because the data set was smaller than at high declinations.  Although
some residual rings due to the $1/f$ noise were evident in the daily
maps (see previous section), these have cleared up in the combined
maps.

\begin{figure}
\vspace{10cm}
\caption{The stacked COSMOSOMAS maps at 13, 15 and 17 GHz.}
\label{Fig:8}
\end{figure}

The Galactic plane crossings, the calibration sources Tau A and Cyg A,
and some of the strongest radio sources can be clearly seen.  Besides
the principal calibrators, Tau A and Cyg A, the expected amplitude of
the strongest radio sources have been estimated from the measurements
by the Michigan monitoring programme (Aller \& Aller, private
communication); these have been complemented by the K\"uhr et al.
(1981) and Green Bank (Condon, Broderick \& Seielstad 1989) catalogues
of discrete radio sources. Away from the Galactic plane the main
contributors are the radio sources 3C 84, 3C 345, 4C 39.25 and 3C 286.
In Table \ref{radio} we present the temperatures measured in the
stacked maps for these sources which are in good agreement with
expectations from the above mentioned catalogues. The uncertainties in
the estimation of these amplitudes are $\sim 450$ $\mu$K at 13 and 15
GHz, and $\sim 800$ $\mu$K at 17 GHz.  Although there are evidence of
the presence of the 3C~286 source in the map at 17 GHz, due to its
small amplitude as compared with the noise in this map, we decided to
quote a 95 \% upper limit.

\begin{table}
\caption{Temperatures  measured for the strongest radio sources ($|b|\ge 10$ degrees) in the COSMOSOMAS maps}\label{radiosources}
\begin{center}
\begin{tabular}{|l|l|l|c|c|c}\hline
Name & $\alpha$ (J2000)&$\delta$ (J2000)& $T_{13}   $ &  $T_{15} $& $T_{17}$ \\

 & & & & (mK) & \\

\hline
3C 84       & $03^{\rm h}19^{\rm m}48^{\rm s}$  &   $+41^{\circ}30'42''$ & 9.4 & 7.2 &  5.5     \\
4C 39.25      & $09^{\rm h}27^{\rm m}03^{\rm s}$  &   $+39^{\circ}02'21''$ & 4.5  &  3.6 &  3.2     \\
3C 286      & $13^{\rm h}31^{\rm m}09^s$    &   $+30^{\circ}31'48''$ &1.5  &  1.4 &   $\le 1.6$     \\
3C345    & $16^{\rm h}42^{\rm m}59^{\rm s}$  &   $+39^{\circ}48'37''$ & 4.9 &  4.1 &  3.3    \\
\\
\end{tabular}
\end{center}
\label{radio}
\end{table}

In the region of the Galactic plane several point-like and extended
sources are detected at the three frequencies. Good agreement between
the positions and fluxes of these sources have been found when comparing
our data with the low-frequency maps at 408 MHz (Haslam et al.
(1982) and 1420 MHz (Reich 1982; Reich \& Reich 1986). This is
illustrated in Fig. 9, where we display these surveys and our data in
the region of Cyg X.  Cyg A is the source on the left of the plot,
while most of the other structure corresponds to the Cyg X complex. A
detailed analysis of the structure detected in our maps and a
comparison between these datasets and the Galactic plane survey at
8.35 and 14.35 GHz (Langston et al. 2000) will be presented in a
forthcoming paper.

\begin{figure}
\vspace{9cm}
\caption{A comparison of the Cyg X region in
the COSMOSOMAS maps and  existing maps at 420 and  1420 MHz.}
\label{Fig:9}
\end{figure}

\section{Conclusions}

1) We have presented a new ground-based CMB experiment working at 13,
15 and 17 GHz and an angular resolution of 1 degree based on a circular
scanning strategy. The performance of the system and the reliability of
the data obtained during the first months of commissioning and operation
have been discussed. We have demonstrated the possibility of removing the
atmospheric and differential ground pickup effectively by the use of
simple techniques that take into account the variation on angular and
time scales of such components.  Daily maps covering ~6000 square
degrees of the sky are routinely obtained  with sensitivities of
$\sim800$ $\mu$K per beam area (FWHM $\sim $ 1\degg) at each frequency.

2) Observations at two different elevations have been performed
providing a stacked map of 9000 square degrees at 13, 15 and 17 GHz 
with mean sensitivities of 140, 150 and 250 $\mu$K per
beam area respectively. These stacked maps show no evidence of
systematics or striping due to $1/f$ noise or residual atmospheric
fluctuations.

3) The strongest radio sources at high Galactic latitudes have been
detected at the levels expected and the structure seen in the Galactic
plane is in good agreement with the low-frequency surveys at 408 and
1420 MHz.

4) Several improvements of the system have been discussed. They include
updating of the filters and optics, which will improve the sensitivity of
the daily data by a factor 2 and will allow a sensitivity of
30 $\mu$K to be achieved after two years of operation.

\section*{Acknowledgements}

We wish to thank especially R. D. Davies and are also grateful to R. J.
Davis, J. Delabrouille, S. Hidalgo, J.  Kaplan, P. Leahy, L.
Piccirillo, B. Revenu and F. Villa for useful suggestions on this
experiment. We are indebted to the mechanical and electronic personnel
of the IAC working at Teide Observatory, who have collaborated in the
installation and operation of the experiment.  J.  Gallegos was
supported by an AECI PhD grant and J. Mac\'\i as-P\'erez acknowledges a
PPARC-(IAC) research studentship.  The COSMOSOMAS experiment is
supported by the Spanish grants DGES PB95-1132-C02-01 and PB
98-0531-C02-02.


\bsp

\label{lastpage}


\begin{thebibliography}{}

\bibitem[]{} Baars J. W. M., Genzel R., Pauliny-Toth I. I. K., Witzel
A., 1977, A\&A, 61, 99

\bibitem[]{} Bennett C. L. et al., 1996, ApJ, 464, L1
 
\bibitem[]{} Condon J. J.,  Broderick, J. J., Seielstad, G. A.,
 1989, AJ, 97, 1064

\bibitem[]{} de Bernardis P. et al., 2000, Nat, 404, 955

\bibitem[]{} de Oliveira-Costa A.  et al., 1999, ApJ, 527, L9

\bibitem[]{} de Oliveira-Costa A. et al., 2001, astro-ph/0010527

\bibitem[]{} Dicker S. R.  et al.,  1999, MNRAS, 309, 750

\bibitem[]{} Davies  R. D. et al.,  1996, MNRAS, 278, 883

\bibitem[]{} Draine B. T., Lazarian A., 1998, ApJ, 494, L19

\bibitem[]{} Gallegos J. E., Gutierrez, C. M., Rebolo, R., Hoyland, R.
J., \& Watson, R. A.  2000, Proceedings of the 19th Texas Symposium on
Relativistic Astrophysics and Cosmology, . Eds.: E. Aubourg, T.
Montmerle, J. Paul, \& P. Peter. NH

\bibitem[]{} Guti\'errez C. M., Rebolo R., Watson R. A., Davies R.
D., Jones A. W., Lasenby A. N., 2000, ApJ, 529, 47

\bibitem[]{} Halverson, N. W. et al. astro-ph/0104489

\bibitem[]{} Hanany, S. et al.,  2000, ApJ, 545, L5

\bibitem[]{} Harrison D. L. et al., 2000, MNRAS, 316, 24

\bibitem[]{} Haslam C. G. T., Salter C. J., Stoffel H., Wilson W.  E.,
1982, A\&AS, 47, 1

\bibitem[]{} Kogut A. et al., 1996a, ApJ, 460, 1

\bibitem[]{} Kogut A. et al., 1996b, ApJ, 464, L5

\bibitem[]{} Kuhr H., Witzel A., Pauliny-Toth L. I. K., Nauber
U., 1981, A\&AS, 45, 367

\bibitem[]{} Langston G., Minter A., D'Addario L., Eberhardt K.,
Koshi K., Zuber J., 2000, AJ, 119, 280

\bibitem[]{} Mauskopf, P. D. et al. 2000, ApJ, 536, L59

\bibitem[]{} Mukherjee P., Hobson M. P.,  Lasenby A. N. 2001, MNRAS, 320, 224

\bibitem[]{} Reich W., 1982,  A\&AS, 48, 219

\bibitem[]{} Reich P.,  Reich, W., 1986, A\&AS, 63, 205

\end{thebibliography}
\end{document}